\documentclass[prl,aps,twocolumn,superscriptaddress,floatfix]{revtex4-2}
\usepackage{graphicx}
\usepackage{times}
\usepackage{color}
\usepackage[version=4]{mhchem}
\definecolor{darkblue}{rgb}{0,0.02,0.45}
\usepackage[colorlinks=true,citecolor=darkblue]{hyperref}
\usepackage{multirow}
\usepackage{comment}

\begin{document}

\title{Magnetism and magnetoelastic effect in 2D van der Waals multiferroic CuCrP$_2$S$_6$}
\author{Jiasen Guo}
\affiliation{Neutron Scattering Division, Oak Ridge National Laboratory, Oak
Ridge, TN 37831, USA}
\author{Ryan P. Siebenaller}
\affiliation{Materials and Manufacturing Directorate, Air Force Research
Laboratory, Wright-Patterson Air Force Base, OH 45433, USA}
\affiliation{Department of Materials Science and Engineering, The Ohio State
University, Columbus, OH 43210 USA}
\author{Michael A. Susner}
\affiliation{Materials and Manufacturing Directorate, Air Force Research
Laboratory, Wright-Patterson Air Force Base, OH 45433, USA}
\author{Jiaqiang Yan}
\affiliation{Materials Science and Technology Division, Oak Ridge National
Laboratory, Oak Ridge, TN 37831, USA}
\author{Zachary Morgan}
\affiliation{Neutron Scattering Division, Oak Ridge National Laboratory, Oak
Ridge, TN 37831, USA}
\author{Feng Ye}
\affiliation{Neutron Scattering Division, Oak Ridge National Laboratory, Oak
Ridge, TN 37831, USA}

\date{\today}

\begin{abstract}
We report a magnetic and neutron diffraction study on the ground state magnetism
and field evolution of single crystal van der Waals multiferroic CuCrP$_2$S$_6$. 
The ordered moments align along the $b$ axis in
the ``A-type'' antiferromagnetic configuration with a spin-flop transition along the
same direction. Field application along $a$ introduces a smooth transition to a
fully-polarized ferromagnetic state via in-plane spin rotation. These findings
resolve the ambiguity of the ground state magnetization direction in
CuCrP$_2$S$_6$ and uncover its field responses, providing a firm basis for
future magnetoelectric study. A magnetoelastic coupling effect connecting the
interlayer spacing and the magnetic order was further revealed, highlighting the
out-of-plane strain as an effective control knob for tuning magnetism both in this
system and in related van der Waals magnets.
\end{abstract}

\maketitle

\section{Introduction}

Ever since the discovery of intrinsic two-dimensional ($2D$) magnetism in
monolayer FePS$_3$~\cite{Wang2016,Lee2016}, and CrI$_3$~\cite{Huang2017}, the
exploration of magnetism in van der Waals (vdW) layered materials 
has become a central focus in advancing spintronic 
technologies~\cite{Blei2021,Liu2023}. Within this context, metal
thio/seleno-phosphates with a general formula MPX$_3$ (M = Fe, Mn, Ni,
X = S, Se) have attracted extensive attention as a versatile material platform,
owing to its chemically tunable magnetic interactions~\cite{Susner2017,
Samal2021}. Introducing heterometallic cations enriches this material family
with polar, electronic, and charge degrees of freedom in addition to magnetism,
giving rise to diverse functional behaviors including ferroelectric
switching~\cite{Liu2016}, optoelectronic response~\cite{Mushtaq2023,Susner2024},
anisotropic electrical conductance~\cite{Wang2023,Sun2025}. The interplay among 
these degrees of freedom further promotes 
multifunctionality~\cite{Harchol2024,Luo2025}. CuCrP$_2$S$_6$ (CCPS) is
a particularly notable example, exhibiting coexisting antiferroelectricity (AFE) and
antiferromagnetism (AFM) with remarkable polarization-magnetization
coupling~\cite{Qi2018,Lai2019, Park2022, Wang2023, Shunta2024, Hu2024}.   

CCPS crystallizes in the monoclinic structure (space group $C2/c$) at ambient
condition~\cite{Maisonneuve1993}, with lamellar layers of edge-sharing CrS$_6$,
P$_2$S$_6$ and CuS$_6$ octahedra that form interconnected triangular
lattices~[Fig.~\ref{Fig1}(a)]. Cr$^{3+}$ ($S$ = 3/2) locates at the
octahedra center, carrying an effective magnetic moment of $\sim 3.8$ Bohr
magneton ($\mu_{\rm B}$). Early magnetization and powder neutron diffraction
studies identified an ``A-type'' AFM order below $T_{\rm N}\sim$~31 K, consisting of
ferromagnetically aligned basal layers stacked antiferromagnetically along the
layer normal direction~\cite{Colombet1982, Cajipea1996}.
In contrast, the nonmagnetic Cu$^+$ resides in an effective double-well
potential~\cite{Colombet1982, Maisonneuve1993,Maisonneuve1997} arising from the
pseudo Jahn--Teller effect~\cite{Burdett1992,Wei1993, Maisonneuve1997,
Fagot-Revurat_2003}. This bistable configuration of CuS$_6$ gives rise to
electric-field-tunable polarization~\cite{lo2023,Ma2023}, offering potential for
memory devices. More importantly, the magnetoelectric coupling (MEC) effect
between the polar order (Cu$^+$) and magnetic order (Cr$^{3+}$) provides a
natural setting for multiferroicity, which has prompted extensive theoretical
and experimental investigations~\cite{Lai2019, Hu2024, Shunta2024}.  

Recently, we elucidated the microscopic ordering mechanism of the
Cu$^+$ sublattice in CCPS~\cite{Guo2026}, thereby establishing a foundation
for investigating MEC in this material. Nevertheless, a definitive
characterization of the magnetic ground state remains elusive.
Early work suggested the ordered moments are oriented along the in-plane
diagonal direction~\cite{Colombet1982, Cajipea1996}, whereas more recent reports
present conflicting assignments, placing the ordered Cr$^{3+}$ spins along the
$a$~\cite{Park2022, Wang2023}, the $b$ axes~\cite{Abraham2025}, or
leaving the moment orientation unspecified~\cite{ Lai2019, Selter2023,
Shunta2024, Luo2025}. Furthermore, the direction-dependent evolution of the
magnetic structure under applied field remains unresolved. This lack of
consensus on the magnetic ground state, together with the absence of studies on
its field response, presents a fundamental barrier to understanding both the
intrinsic magnetism of CCPS and its proposed multiferroic behavior.

In this article, we present a comprehensive neutron diffraction and
magnetization study of the magnetic ground state of CCPS and its field
response. We identify a low temperature ``A-type" AFM spin configuration with
magnetic easy axis along the $b$ axis. A spin-flop transition occurs under field
applied along the same direction. In contrast, magnetic field applied along $a$ 
drives a continuous spin rotation towards a fully-polarized ferromagnetic (FM) state.
A magnetoelastic coupling between the spin canting angle and the interlayer
spacing is also revealed, suggesting a new route for controlling magnetism
through out-of-plane strain in this vdW antiferromagnet.

\section{Experimental details}
Single crystals CCPS were synthesized from pure elements following the procedure
described in Ref.~\cite{Susner2020}. Magnetic measurements were performed
using a Quantum Design magnetic property measurement system (MPMS) on a $\sim$
0.4 mg crystal, with the crystallographic directions determined by x-ray Laue
diffraction. Single crystal x-ray measurements were carried out on a CCPS flake
with dimensions of $\rm 5\times80\times80~\mu m^3$ mounted to a Mitegen loop
using Paratone-N oil. Data were collected on a Rigaku Synergy-DW diffractometer
which uses Mo source with wavelength of 0.7107~\AA. The sample temperature is 
regulated using Oxford cryostream. The reduced data were processed with the
CrysAlisPro software \cite{crysalispro}. Neutron diffraction measurements were performed on the
time-of-flight single crystal diffractometer CORELLI at the Spallation Neutron
Source (SNS) at Oak Ridge National Laboratory (ORNL)~\cite{ye18}. 
22 pieces of single crystals (mass $\sim$ 8~mg) were co-aligned on an aluminum plate
mounted to a sample stick. The sample mosaic was estimated to be less than
3$^\circ$. For the measurements in zero field, a closed-cycle refrigerator was
used with a base temperature of $\sim$ 4~K. For field-dependent measurements, 
a 5-Tesla superconducting cryro-magnet was used. The reciprocal space was surveyed by
rotating the sample stick 360$^{\circ}$ at 3$^\circ$ steps at each temperature
and field. The temperature and field dependence of the magnetism was
characterized via monitoring the magnetic peak intensity at constant ramping
rate.

\section{Results}
\begin{figure}[htp]
  \includegraphics[width=\linewidth]{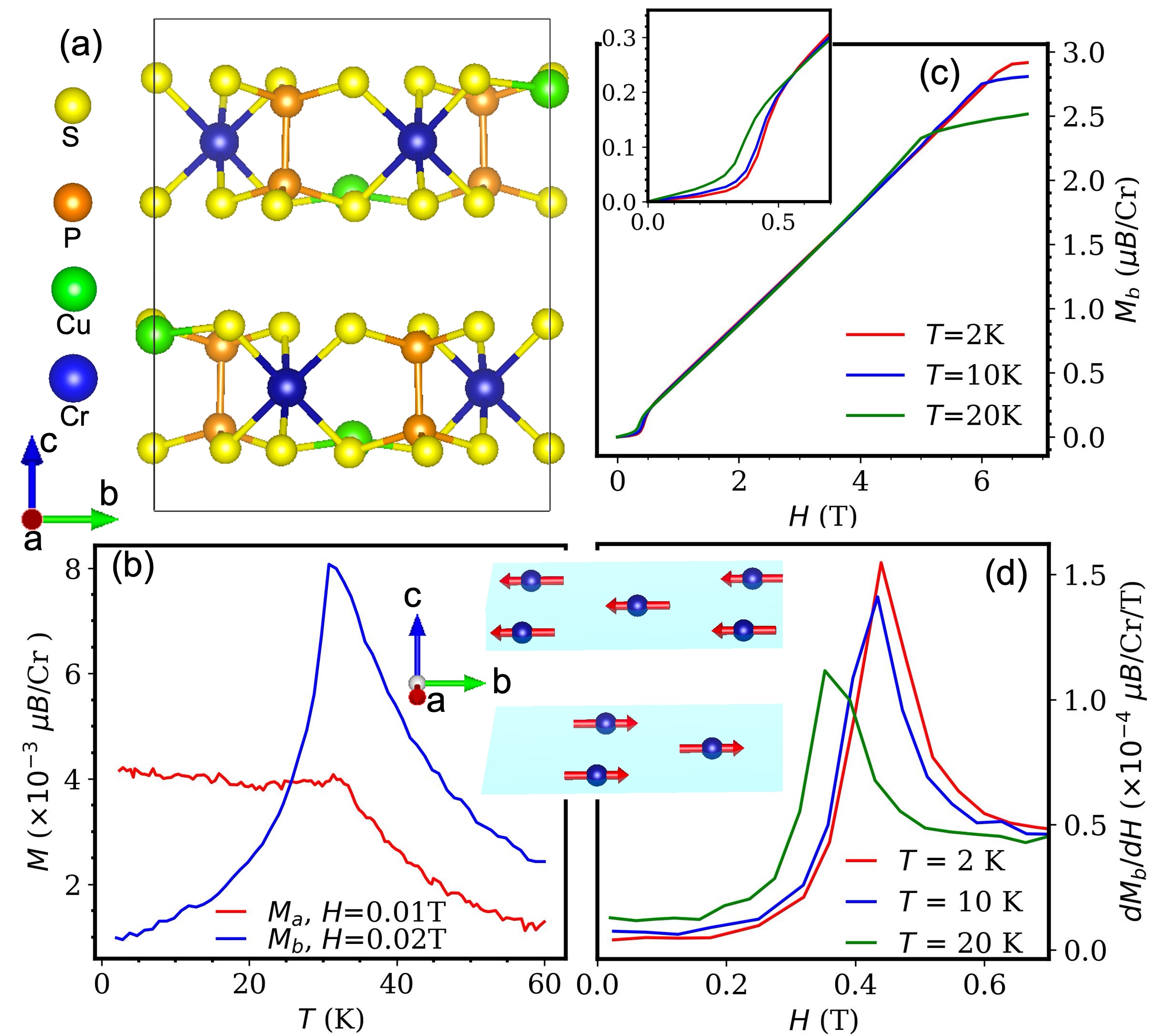}
  \caption{(a) Crystal structure of CuCrP$_2$S$_6$ in the AFE
  state. (b) Field cool magnetization $M_a$ and $M_b$ along the $a$ and $b$
  crystallographic axes, respectively. Inset shows the refined magnetic 
  structure. (c) Field dependence of $M_b$ at
  selective temperatures. Inset highlights the spin-flop transition in the low
  field regime. (d) Derivative of the $M_b$ with respect to the field strength.
  The peak position defines the spin-flop transition critical field $H_{\rm
  SF}$. }
  \label{Fig1}
\end{figure}

Figure~\ref{Fig1}(b) presents the temperature dependence of $M_a$ and $M_b$ near
the magnetic transition. Both increase as the temperature approaches $T_{\rm N}$
from above. At $T_{\rm N}$ = 31.2 K, $M_b$ exhibits a sharp $\lambda$-like cusp,
whereas $M_a$ maintains nearly constant below $T_{\rm N}$. These features are
characteristic of an AFM transition with moments along the $b$
direction~\cite{Kittel1971}, consistent with Ref.~\cite{Abraham2025}.
Fig.~\ref{Fig1}(c) shows the field dependence of $M_b$ at various temperatures.
At $T$ = 2 K, $M_b$ displays a low-field anomaly, followed by a linear trend
and eventual saturation at $H$ = 6.5 T. Such anomalies are commonly associated
with a spin-flop transition~\cite{Abraham2025, wang2025, Park2022}, in which a
magnetic field applied along the N{\'e}el vector induces its rotation to
minimize the Zeeman energy. This observation further supports that the ordered
moments are aligned along the $b$ direction in the ground state. As $T$
approaches $T_{\rm N}$, the saturation field decreases gradually, and the
spin-flop transition becomes increasingly broadened.

 The critical field $H_{\rm SF}$, defined as the peak position in the derivative
 of $M_b$ with respect to $H$ ($dM_b/dH$), is determined to be 0.44 T at $T$ = 2 K
 [Fig.~\ref{Fig1}(d)], in good agreement with
 Ref.~\cite{Abraham2025}. It decreases to $H_{\rm SF}$ = 0.43 T and 0.35 T at
 $T$ = 10 K and 20 K, respectively, and is expected to approach zero at $T_{\rm
 N}$ as the magnetic order vanishes. The saturation field $H_{\rm
 sat}$ also decreases with increasing temperature, reflecting a progressive 
 weakening of the exchange interactions due to thermal fluctuations. Because the
 intralayer FM interaction is dominant in CCPS~\cite{Colombet1982,
 Abraham2025}, its magnetic structure can be approximated as a stacking 
 of rigid ferromagnetic layers coupled antiferromagnetically within a 
 mean-field framework. This description
 enables extraction of the effective anisotropy field $H_{\rm A}$ and
 exchange field $H_{\rm E}$ using the relations $H_{\rm SF} = \sqrt{2H_{\rm E}H_{\rm A}-H^2_{\rm
 A}}$ and $H_{\rm sat} = 2H_{\rm E}-H_{\rm A}$~\cite{Jaccarino1983}. These 
 parameters characterize the magnetic anisotropy within individual layers
 and the interlayer exchange stiffness. They are related to the single ion
 anisotropy constant $K$ and interlayer AFM exchange coupling $J_c$ through $H_{\rm
 {A}}=2KS/g\mu_{\rm B}$ and $H_{\rm {E}}=2J_cS/g\mu_{\rm B}$. The extracted 
 values are summarized in Table~\ref{tab:estimation}, which enables modeling of
 $M_a$~\cite{supp_mat} and determination of the field dependence of the canting angle
 between the adjacent FM layers, to be discussed in Fig.~\ref{Fig3}(b).

\begin{table}[bht]
\caption{Extracted exchange field $H_{\rm E}$, anisotropy field $H_{\rm A}$, and
saturation magnetic moments $M_{\rm s}$ from the magnetization data at different
temperatures based on the stacking model with AFM-coupled rigid FM layers.}
\begin{ruledtabular}
\begin{tabular}{cccc}
$T$ (K)&$H_{\rm E}$ (T) &$H_{\rm A}$ (T)&$M_{\rm s}$ $(\mu_{\rm B}$/Cr$^{3+}$)\\
\hline
2 & 3.36 & 0.03 & 2.90\\
10 & 3.02 & 0.03 & 2.75\\
20 & 2.51 & 0.02 & 2.33\\
\end{tabular}
\label{tab:estimation}
\end{ruledtabular}
\end{table}

\begin{figure}[htp]
  \includegraphics[width=\linewidth]{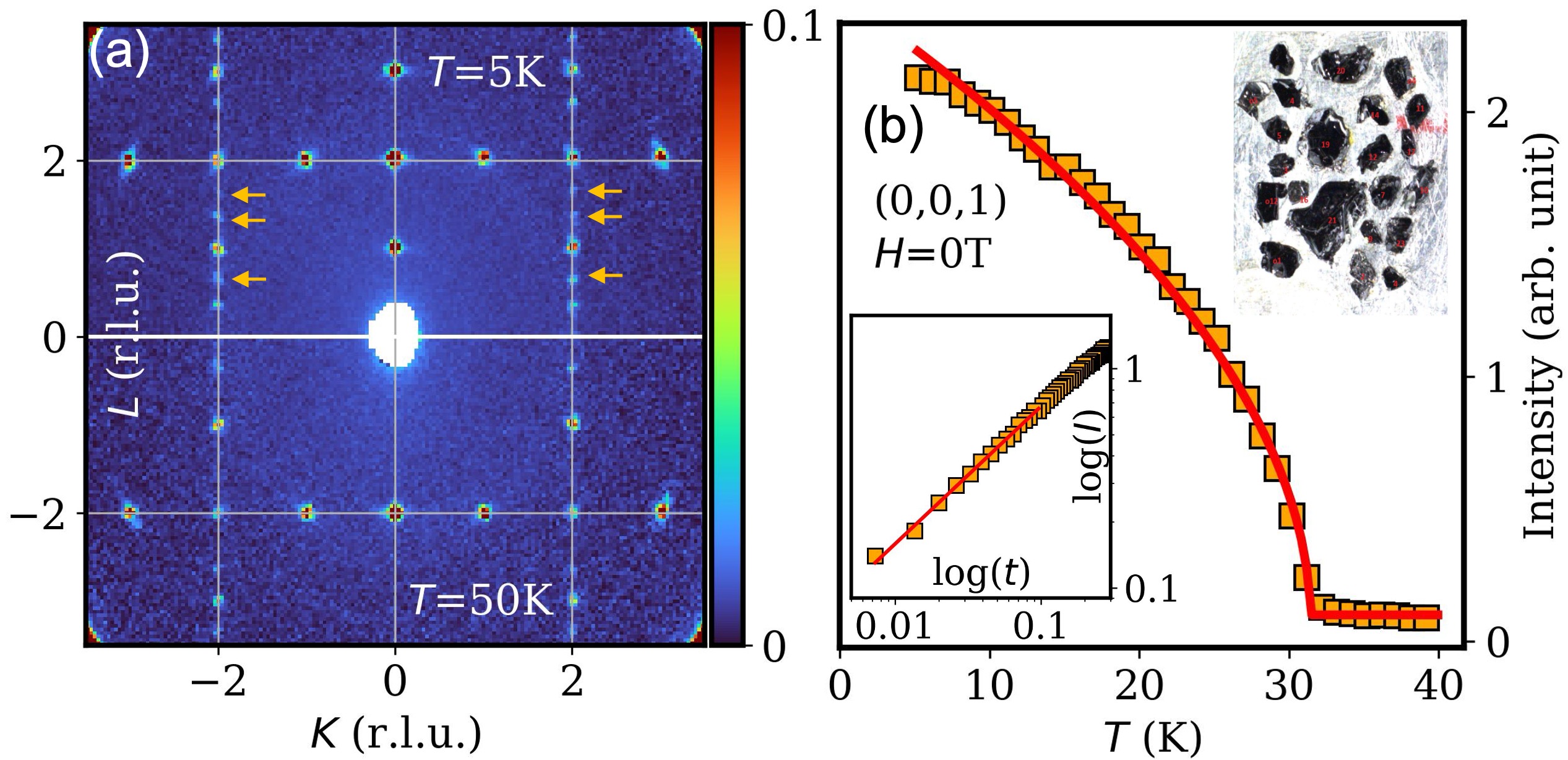}
  \caption{(a) Neutron diffraction pattern in the $(0KL)$ scattering plane at $T
  = 5$~K (top) and $T = 50$~K (bottom)  at zero field. Orange arrows mark the
  magnetic peaks from other structural domains. (b) The temperature dependence
  of the (0, 0, 1) magnetic peak intensity at zero field. The solid line is a
  power law fit of $I\propto|t|^{2\beta}$ with the reduced temperature
  $t=(T_{\rm N}-T)/T_{\rm N}$. The upper inset shows the co-aligned single
  crystal array used in the neutron measurements. The lower inset shows the
  $\log-\log$ plot with a linear fit in the range $|t|<0.1$ to determine
  $\beta$.} 
  \label{Fig2}
\end{figure}

Neutron diffraction further confirms the direction of the ordered magnetic
moment. Fig.~\ref{Fig2}(a) presents the diffraction patterns in the $(0KL)$
plane at 5 and 50~K. In addition to the nuclear Bragg peaks at integer
positions, weak fractional peaks appear at $L$ = $1/3$ and $2/3$, arising from
twinned structural domains associated with the pseudo-hexagonal symmetry in the
basal plane~\cite{supp_mat}. At $T$ = 5~K, strong magnetic reflection appear at
(0, 0, 2$n$+1) positions indicating AFM coupling between adjacent layers.
Fractional magnetic peaks originating from other structural domains are also
present (marked by arrows). The temperature dependence of the (0, 0, 1) peak
intensity yields $T_{\rm N}$ = 31.4(2)~K, in good agreement with the
magnetization data [Fig.~\ref{Fig2}(b)]. The extracted component
$\beta$ = 0.314(5) suggests a $3D$ critical behavior near the transition. Together
with the absence of magnetic diffuse scattering (indicating negligible critical
fluctuation), the presence of significant interlayer magnetic interactions, and
weak in-plane anisotropy, demonstrate that CCPS is an anisotropic $3D$ magnetic
system. 

Detailed magnetic structure refinement based on representational analysis reveals an
``A-type" AFM configuration (magnetic space group $Pc^\prime$), consisting of FM layers
within the $ab$ plane that stack antiferromagnetically along the layer normal~\cite{supp_mat}. 
The ordered moments of 3.03(5) $\mu_{\rm B}$, align along the $b$ direction, 
in agreement with the magnetization results [Fig.~\ref{Fig1}(b)]. 
Notably, the absence of magnetic scattering at (0, 2, 0) position provides 
additional support for this moment orientation, since the magnetic neutron scattering 
is sensitive only to the spin component perpendicular to the momentum transfer,
[$\mathbf{Q}\times (\mathbf{M} \times \mathbf{Q}$)]. 

To characterize the field evolution of magnetism in CCPS, neutron diffraction
measurements were carried out under a magnetic field applied along the $a$
direction (w.r.t.~the main structural domain). Fig.~\ref{Fig3}(a) shows the
field dependence of the (0, 0, 1) and (0, 2, 0) peak intensities at $T$ = 27~K.
As the field increases, the intensity of the (0, 0, 1) peak decreases progressively 
and vanishes above a critical field $H_{\rm FM}$ = 3.7~T. Concurrently, the
(0, 2, 0) peak emerges and gradually saturates near the same
field. The zero-field intensity is fully recovered upon removing the field,
implying negligible hysteresis. Upon cooling, $H_{\rm FM}$
increases to approximately 4.5~T at $T$ = 22~K and quickly exceeds the
accessible field range~\cite{supp_mat}.

This change in magnetic intensity can be understood as a continuous,
symmetry-adapted evolution from the AFM configuration to a
field-polarized FM state with moments aligned along $a$~[Fig.~\ref{Fig3}(a) inset].
To quantify this evolution, we adopt the same simplified stacking model
used in the analysis of the magnetization data and employed an empirical
relation, $\theta \propto AH^2+BH+C$, to describe the canting angle $\theta$
between the FM moment and the $b$ direction as a function of magnetic field $H$, where
$A, B, C$ are constants~\cite{supp_mat}. By calculating the magnetic structure
factor at a given $\theta$, the (0, 0, 1) magnetic peak intensity can be
fitted, allowing $\theta$ to be extracted as a function of $H$. Fig.~\ref{Fig3}(b)
plots the resulting $\theta$ values at selected temperatures and fields. Below
$T$ = 10~K, $\theta$ remains below 40$^\circ$ up to $\sim$ 4~T, indicating
robust interlayer AFM coupling against thermal fluctuation. As the temperature
increases, the field response of $\theta$ becomes progressively softer. Full
polarization is reached at $\sim 3.7$~T at $T$ = 27~K, consistent with the
disappearance of the (0, 0, 1) peak intensity and the saturated (0, 2, 0) peak.

Energetic analysis within the stacking model shows that the canting angle
$\theta$ can be determined from the relation $\sin\theta=H/(2H_{\rm E} + H_{\rm
A})$ for a magnetic field $H$ applied along the $a$ direction~\cite{supp_mat}.
Using the parameters listed in Table~\ref{tab:estimation}, the calculated field
dependence of $\theta$ agrees well with the values obtained from fitting the
neutron data [Fig.~\ref{Fig3}(b)]. Applying this expression and assuming
negligible anisotropy at $T$ = 27~K yields an estimate of $H_{\rm E}\sim 1.35$
T, highlighting the rapid evolution of exchange coupling as the system approaches
$T_{\rm N}$.
 
\begin{figure}[htp]
  \includegraphics[width=\linewidth]{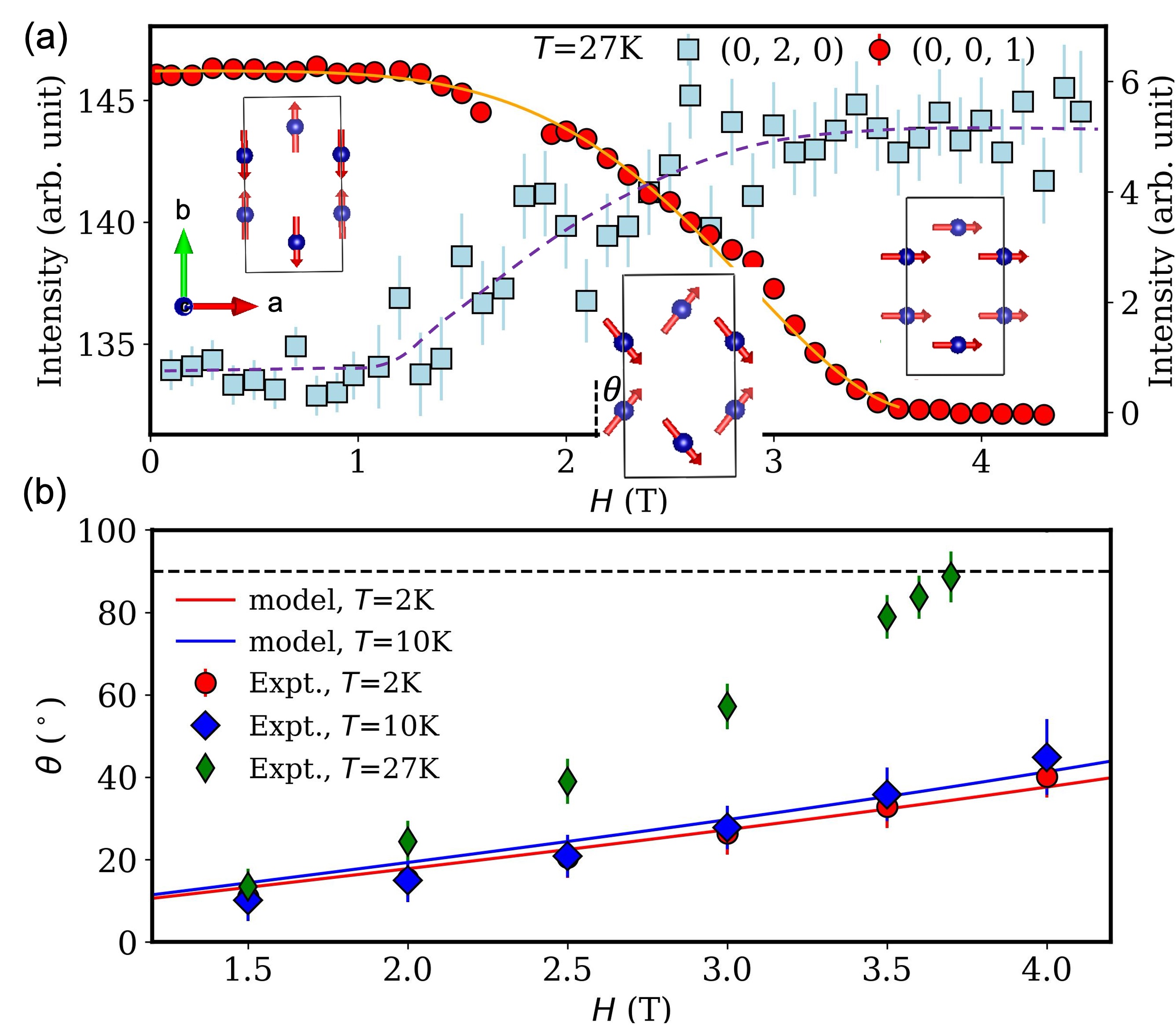}
  \caption{Field evolution of the magnetic structure of CCPS with $a$-axis
  in-plane field. (a) Field dependence of the (0, 0, 1) and (0, 2, 0) peak
  intensities at $T$ = 27 K. Insets illustrate the evolution of the magnetic
  structure with increasing field. The solid line is the model fit and the
  dashed line is guide to the eye. (b) Field dependence of the in-plane spin
  canting angle $\theta$. Symbols are experimental values obtained via fitting
  the (0, 0, 1) peak intensity at different temperatures, solid lines are model
  predicted values using the stacking model. The dashed line marks the
  fully-polarized state with $\theta$ = 90$^\circ$.}
  \label{Fig3}
\end{figure} 

\begin{figure}[htp]
  \includegraphics[width=\linewidth]{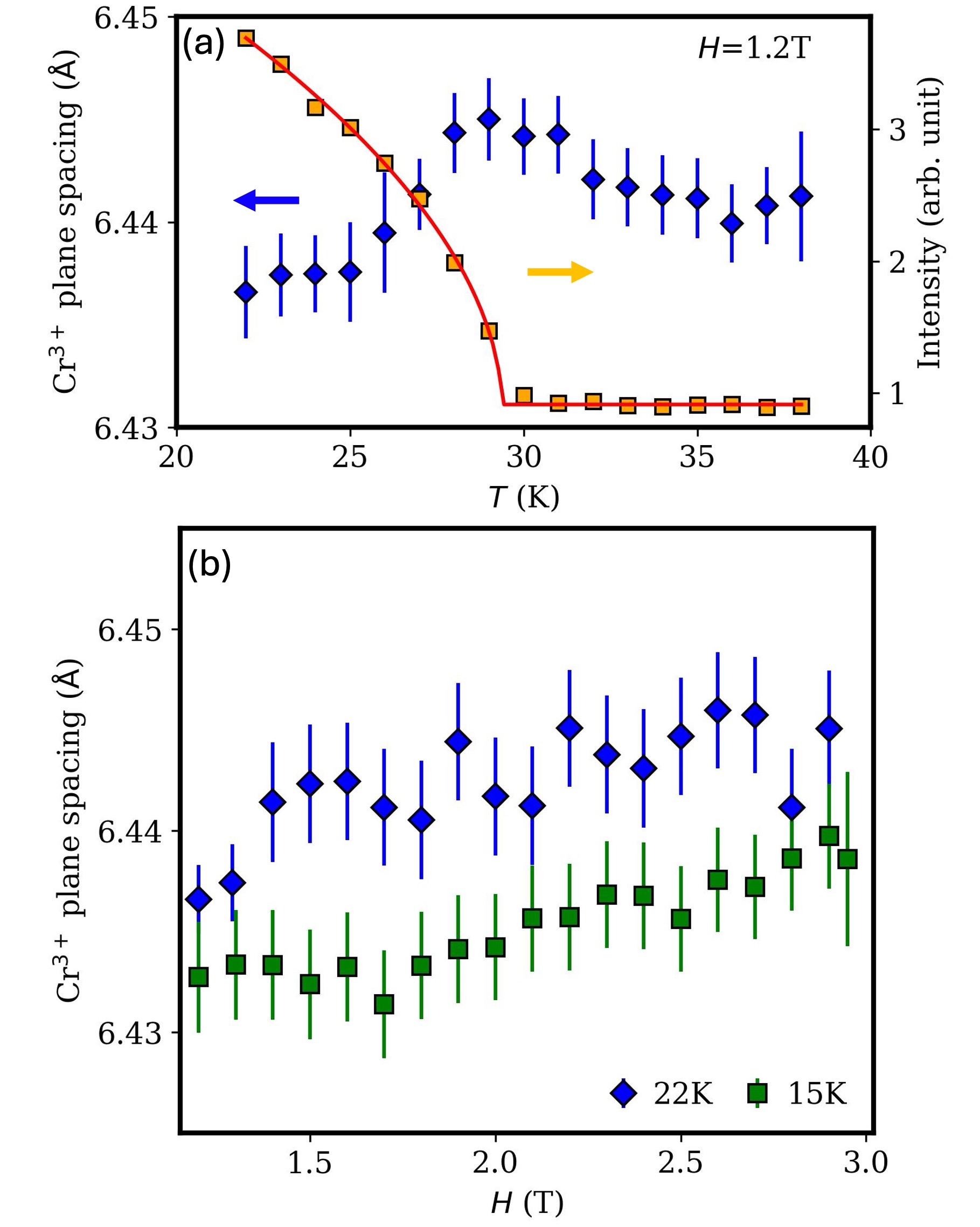}
  \caption{Magnetoelastic coupling effect. (a) Temperature dependence of (0, 0,
  1) magnetic peak intensity and the interlayer spacing under $H$ = 1.2 T along
  the $a$ direction. The solid line is a power law fit. (b) Field evolution of
  the interlayer spacing at $T$ = 15~K and 22~K.}
  \label{Fig4}
\end{figure}

Figure~\ref{Fig4}(a) presents the $T$-dependence of the
(0, 0, 1) magnetic peak intensity at $H$ = 1.2 T along $a$.
Compared to the zero-field case, the AFM transition is suppressed to
$T_{\rm N}$ = 29.4(1)~K. Although a net moment along $a$ is
symmetry-allowed, the reduced $T_{\rm N}$ implies that this configuration is
energetically less favorable, consistent with the magnetic 
easy axis lying along the $b$ direction. 
The onset of the AFM order is accompanied by a broad maximum in the
interlayer spacing between adjacent Cr$^{3+}$ planes. The increase in 
interlayer spacing upon cooling reflects negative thermal expansion (NTE)
along the layer normal direction~\cite{Susner2020}. In contrast, 
the rapid decrease below $T_{\rm N}$ indicates an energy minimization driven by AFM
ordering. The observed hump in the interlayer spacing therefore arises from the
combined effect of NTE and magnetic ordering. 
Below $T_{\rm N}$, applying field along $a$ leads to an expansion of the
interlayer spacing~[Fig.~\ref{Fig4}(b)]. This behavior can be
understood as a competition between the interlayer AFM coupling $J_c$
and the Zeeman energy; increasing field enhances the canting angle,
thereby increasing the net FM moment and imposing an energy penalty on
$J_c$. As a result, the interlayer spacing expands to minimize the total energy.

\begin{figure}[htp]
  \includegraphics[width=\linewidth]{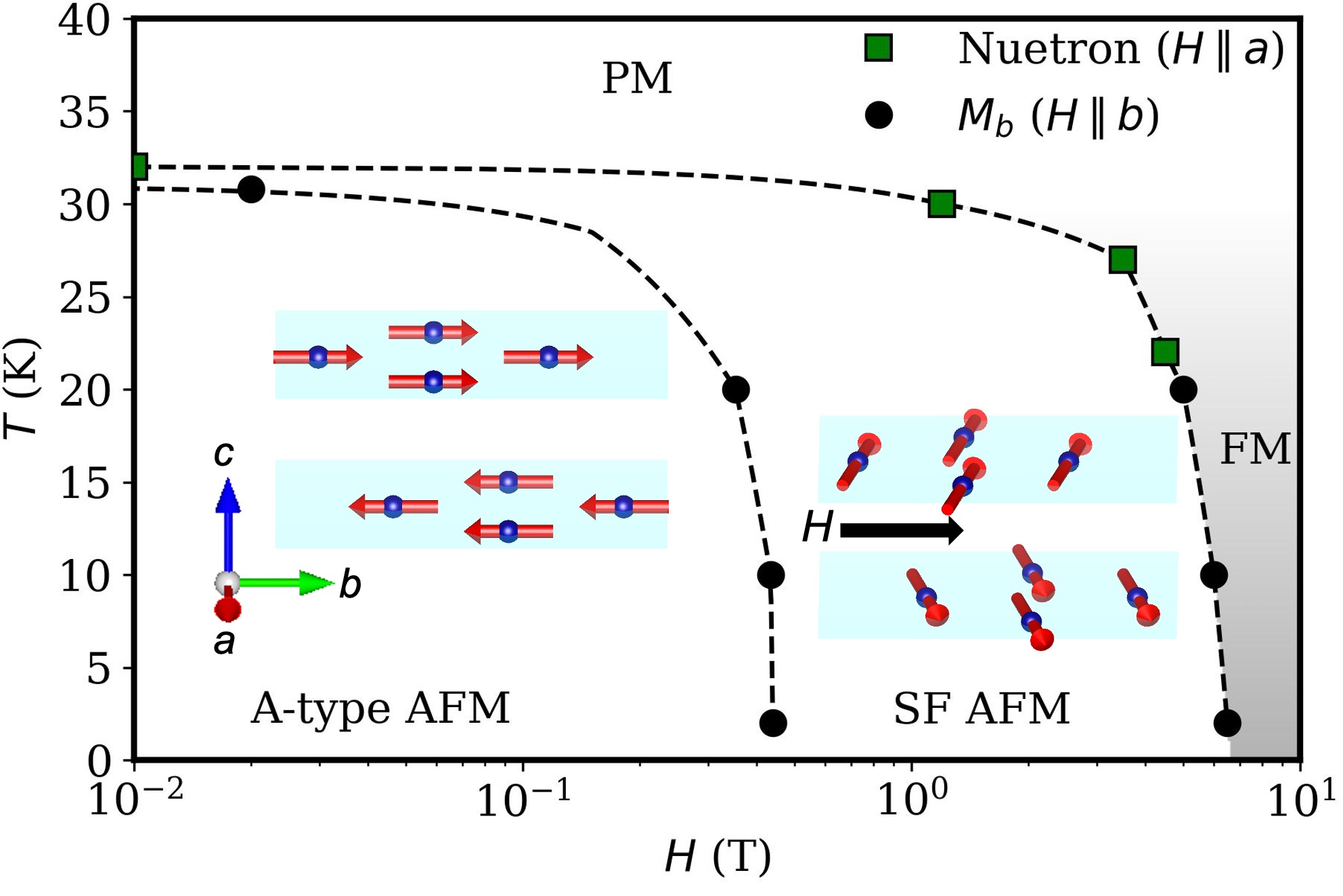}
  \caption{$T$-$H$ phase diagram of the magnetic structure in CCPS. Insets
  illustrate the magnetic structure at $H$ = 0~T, the spin-flop AFM
  configuration at $H_b > H_{\rm SF}$. Colored symbols are obtained from neutron
  diffraction results with field applied along the $a$ direction.}
  \label{Fig5}
\end{figure}

Based on the neutron scattering and magnetization results,
Fig.~\ref{Fig5} summarizes the magnetic phase diagram of CCPS. At
high temperature, the Cr$^{3+}$ sublattice is paramagnetic. Below $T_{\rm N}$ =
31 K, an ``A-type" AFM phase is stabilized, with the ordered moment
aligned along $b$. When a magnetic field is applied along $b$, exceeding
a critical field $H_{\rm SF}$ induces a spin-flop transition, 
in which the staggered moments rotate toward $a$ while retaining a net
component along $b$. In contrast, a field along the $a$ direction drives a 
symmetry-adapted canted AFM phase with a net moment along $a$. In both field
orientations, a sufficiently strong field fully polarizes the Cr$^{3+}$ sublattice.

\section{Discussions}
The pseudo-hexagonal monoclinic metric enables the formation of rotated 
structural domains, a phenomenon well-known in layered
chalcogenides and halides, where twisted stacking gives rise to
fractional diffraction peaks~\cite{Jang2024}. This effect highlights the
difficulty of obtaining reliable single crystal magnetization data and may
account for the lack of consensus regarding the ordered moment direction 
in recent literature.
The misassignment of ordered moment direction in earlier powder neutron
studies likely originates from peak overlap between (0, 2, 0) and (-1, 1, 0) 
reflections, caused by their close scattering angles and the limited 
instrument resolution~\cite{Cajipea1996}. 

The observed magnetic anisotropy is relatively weak, as evidenced by 
the small value of $H_{\rm A}$. In general, magnetic anisotropy arises from 
three primary contributions: shape anisotropy, single-ion anisotropy, 
and exchange anisotropy~\cite{Hou2025}. Although shape anisotropy can be significant 
in 2D materials and may favor in-plane magnetization in CCPS,
it cannot account for the easy-axis alignment along $b$, particularly 
given the irregular sample geometry used in this work. 
Single-ion anisotropy arises from a combined effect of spin-orbit coupling (SOC) and crystal
electric field (CEF). In an approximate trigonal-distorted octahedron
environment, Cr$^{3+}$ adopts a half-filled low energy manifold in the CEF
ground state, similar to that in MnPS$_3$ and NiPS$_3$~\cite{Autieri2022}. The absence
of orbital degeneracy leads to a quenched orbital moment~\cite{Lado2017,
Suzuki2022}, as evidenced by the nearly isotropic Land\'{e} $g$ factor $\sim$
2~\cite{Colombet1982, Abraham2025} and an ordered moment $\sim$ 3~$\mu_{\rm B}$, in
excellent agreement with the classical spin-only limit. This indicates that single ion
anisotropy originates from second order SOC via quantum fluctuations of the
orbital moment~\cite{Lado2017}. Consequently, the anisotropy is
expected to be weak, scaling as $\lambda^2/\Delta$, where $\lambda$ is
the SOC strength for Cr$^{3+}$ and $\Delta$ denotes the energy separation between 
the CEF ground and excited states~\cite{Lado2017}. 

On the other hand, magnetic anisotropy in vdW honeycomb magnets with quenched
orbital moments, such as Cr$_2$Ge$_2$Te$_6$ and CrI$_3$~\cite{Lado2017}, is
known to originate from anisotropic exchange interactions mediated by the SOC of 
ligand anions~\cite{Lado2017, Xu2018}. Anisotropic exchange interactions
through extended pathway (M-O-P-O-M) have also been demonstrated in  
honeycomb systems bridged by PO$_4^{3-}$ units~\cite{Abdeldaim2024}. In CCPS,
the Cr$^{3+}$ sublattice is linked by P$_2$S$_6$ units, forming a comparable 
exchange pathway (Cr-S-P-S-Cr). This suggests that ligand-mediated anisotropic
exchange may play a key role in determining the magnetic anisotropy. 
We therefore encourage further theoretical investigations, particularly DFT calculations
including full SOC effects, to clarify the microscopic origin of 
magnetic anisotropy in this system.

The interlayer spacing in 2D vdW magnets has been identified as a crucial tuning
parameter that controls the interlayer exchange coupling mediated by an M-X-X-M
exchange pathway involving chalcogen $p$-orbital overlap across the vdW
gap~\cite{Wang2020}. A reduced interlayer spacing favors AFM interlayer, as
Pauli repulsion  in the overlap region dominates over the kinetic energy gain
associated with interlayer hopping~\cite{Wang2020}. A similar mechanism,
involving an interlayer Cr-S-S-Cr pathway, has also been proposed to explain the
tuning of $J_c$ in CCPS~\cite{Hu2024}. In this context, the competition between
these energy contributions provided a natural explanation of the observed
reduction in the interlayer spacing below $T_{\rm N}$.

This temperature dependence, together with the field tuning of the interlayer
spacing, emphasizes the presence of magnetoelastic coupling in CCPS. Such coupling 
is widely exploited in straintronics, where strain engineering enables effective control
of a material's magnetic and electronic properties~\cite{Bukharaev2018,
Bandyopadhyay2021}. VdW materials are particularly attractive in this context
due to their exceptional mechanical properties, including anisotropic
compressibility~\cite{Miao2021}. For example, out-of-plane strain has been shown to 
significantly enhance interlayer AFM coupling in the vdW magnet
CrI$_3$~\cite{Song2019, Li2019}. A linear relationship between $T_{\rm N}$ and
interlayer spacing under applied strain has also been reported in
MnPS$_3$~\cite{Toyoshima2009}. Therefore, the magnetoelastic coupling observed
in this study points to a promising route for tuning the 
magnetic properties in CCPS.
Recent studies on the isostructural CuInP$_2$S$_6$
have demonstrated control of polarization via out-of-plane strain, where
local strain gradients drive ferroelectric domain formation through the flexoelectric
effect~\cite{Chen2022, Ming2022}. Such strain gradients are also expected
to modulate the interlayer exchange coupling $J_c$ by altering the layer spacing 
and Cr-S covalent bonding. Accordingly, investigating strain-based control 
mechanism in CCPS presents both significant challenges and exciting opportunities,
owing to the coupled tunability of its polarization and magnetic properties.

\section{Conclusion} 
In summary, we have investigated the magnetic ground state 
of CCPS and its evolution under an applied magnetic fields using
neutron scattering and magnetic measurements. The ordered moments in the
``A-type" AFM structure are found to align along the $b$ axis. A 
spin-flop transition is observed when the field is applied along this 
direction, indicating  weak in-plane uniaxial anisotropy. These results resolve 
the long-standing ambiguity
regarding the magnetization orientation and pave the way for future studies
of MEC mechanism in this system. A magnetic field applied along $a$
stabilizes a symmetry-adapted canted AFM state, accompanied by a
pronounced magnetoelastic coupling effect. This coupling links the spin canting angle 
to the interlayer spacing, highlighting the intricate interplay between lattice and
spin degrees of freedom. Our findings provide general guidelines for exploring similar
phenomena in related 2D vdW multiferroics. Furthermore, out-of-plane strain is expected to
modify the magnetic ordering temperature and spin configuration by tuning
the interlayer spacing. Local strain gradient may enable simultaneous control of
polarization and magnetism through the flexoelectric effect,
opening promising opportunities for engineering multifunctional responses
in low-dimensional materials.

\section{Acknowledgement}
This research  used resources at the Spallation Neutron Source, a DOE Office of
Science User Facility operated by the Oak Ridge National Laboratory. The beam
time was allocated to CORELLI on proposal number IPTS-33430.1. 
Sample synthesis and characterization was performed at the Materials and
Manufacturing Directorate at the Air Force Research Laboratory and were
supported by the United States Air Force Office of Scientific Research (AFOSR)
Grant LRIRs 23RXCOR003 and 26RXCOR010. J.-Q. Y. was supported by the U.S.
Department of Energy, Office of Science, Basic Energy Sciences, Materials
Sciences and Engineering Division.

This manuscript has been authored by UT-Battelle, LLC, under Contract No.
DE-AC0500OR22725 with the U.S. Department of Energy. The United States
Government retains and the publisher, by accepting the article for publication,
acknowledges that the United States Government retains a non-exclusive, paid-up,
irrevocable, world-wide license to publish or reproduce the published form of
this manuscript, or allow others to do so, for the United States Government
purposes. The Department of Energy will provide public access to these results
of federally sponsored research in accordance with the DOE Public Access Plan
(http://energy.gov/ downloads/doe-public-access-plan).

%\bibliographystyle{h-physrev}
%\bibliography{reference.bib}

\end{document}

\title{Supplementary Material: Magnetism and magnetoelastic effect in 2D van der Waals multiferroic CuCrP$_2$S$_6$}
\author{Jiasen Guo}
\affiliation{Neutron Scattering Division, Oak Ridge National Laboratory, Oak Ridge, TN 37831, USA}
\author{Ryan P. Siebenaller}
\affiliation{Materials and Manufacturing Directorate, Air Force Research Laboratory, Wright-Patterson Air Force Base, OH 45433, USA}
\affiliation{Department of Materials Science and Engineering, The Ohio State University, Columbus, OH 43210 USA}
\author{Michael A. Susner}
\affiliation{Materials and Manufacturing Directorate, Air Force Research Laboratory, Wright-Patterson Air Force Base, OH 45433, USA}
\author{Jiaqiang Yan}
\affiliation{Materials Science and Technology Division, Oak Ridge National Laboratory, Oak Ridge, TN 37831, USA}
\author{Zachary Morgan}
%\email{zgf@ornl.gov}
\affiliation{Neutron Scattering Division, Oak Ridge National Laboratory, Oak Ridge, TN 37831, USA}
\author{Feng Ye}
%\email{2fy@ornl.gov}
\affiliation{Neutron Scattering Division, Oak Ridge National Laboratory, Oak Ridge, TN 37831, USA}

\maketitle

\section{Crystal direction determination}
\begin{figure*}[htp]
  \includegraphics[width=0.8\linewidth]{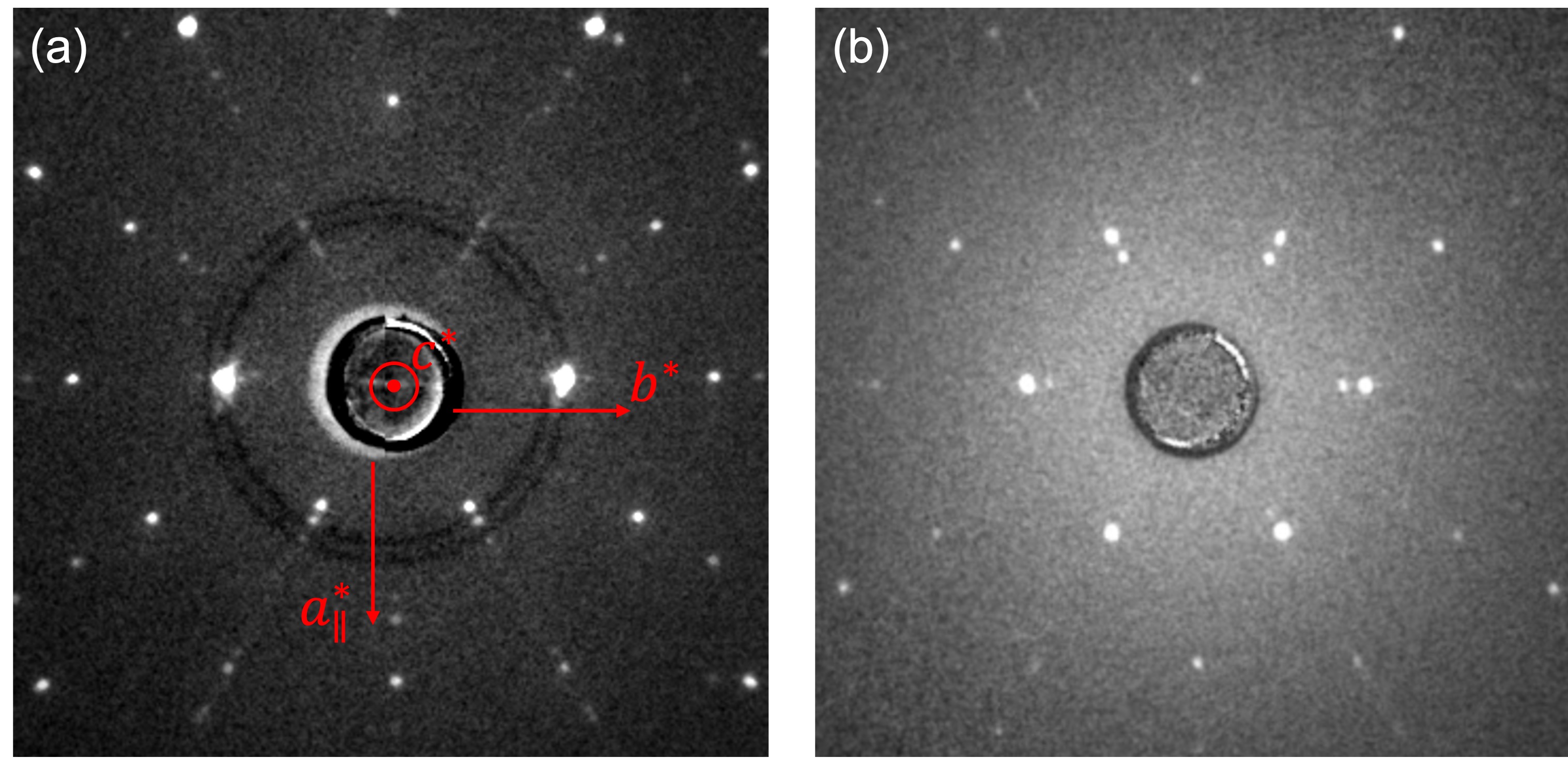}
  \caption{Room temperature X-ray Laue pattern of CCPS with (a) single domain (b) multiple domains. The reciprocal lattice directions are labeled with $a^\ast_\parallel$ denoting the projected component of $a^\ast$ onto the $ab$ plane. }
  \label{FigS1}
\end{figure*}

Fig.~\ref{FigS1}(a) shows the room temperature X-ray Laue pattern of the single crystal CCPS flake used in the magnetization measurements. The two-fold symmetry of the reciprocal space about the $a^\ast c^\ast$ plane is clearly demonstrated for the monoclinic structure (space group $C2/c$) with the $b$-unique axis. In contrast, Fig.~\ref{FigS1}(b) shows a typical X-ray Laue pattern lacking the two-fold symmetry, indicating coexisting structural domains. Such samples are coaligned for the neutron diffraction measurements. 

\clearpage

\section{Simulation of the spin canting angle}
\begin{figure*}[htp]
  \includegraphics[width=0.8\linewidth]{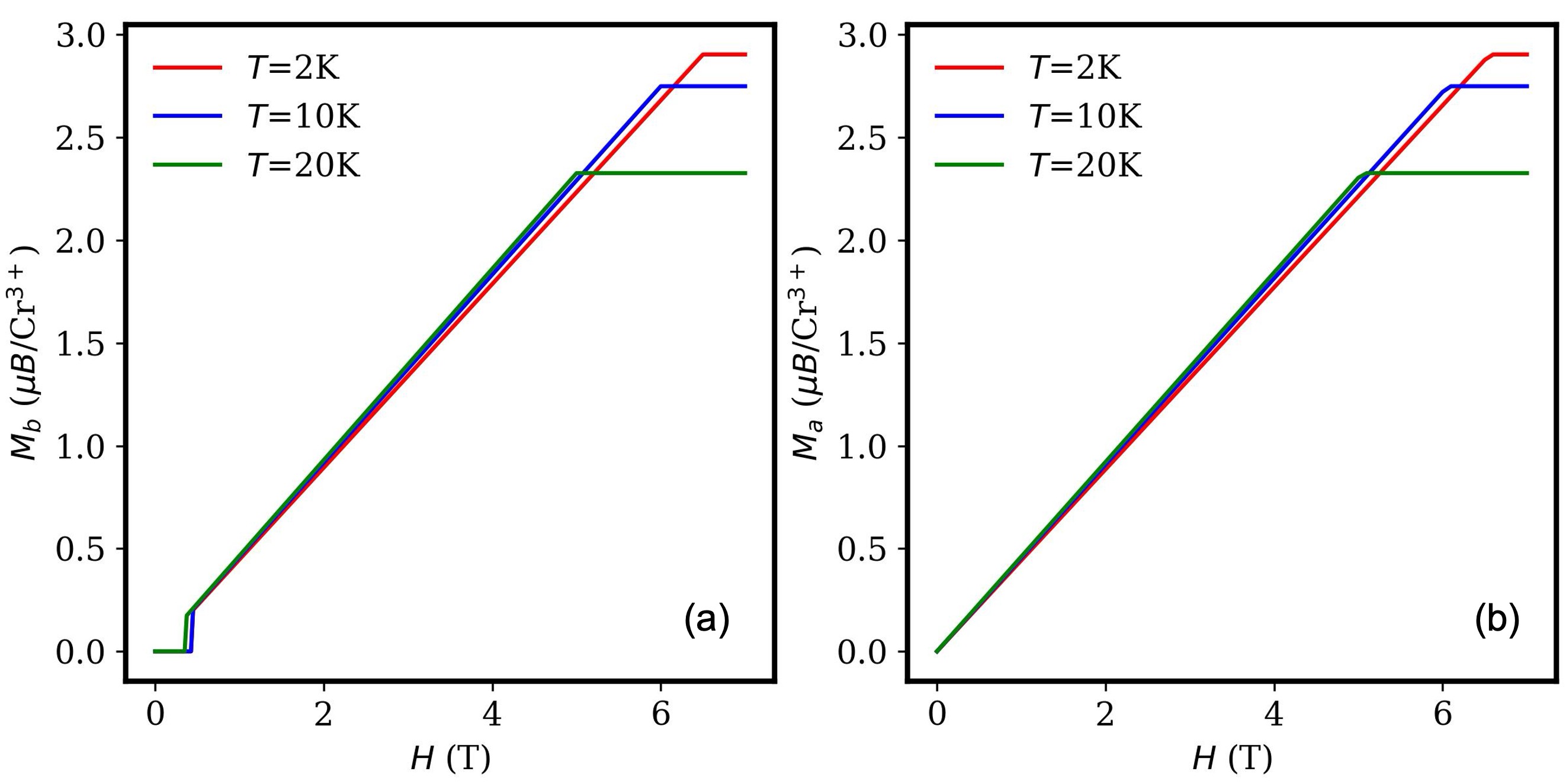}
  \caption{Simulated magnetization (a) $M_b$ and (b) $M_a$, respectively.}
  \label{FigS2}
\end{figure*}

The magnetic structure of CCPS can be approximated with a simplified model consisting AFM-coupled rigid FM layers stacking along the layer normal direction. The atomistic Hamiltonian of such a model is given by
\begin{equation}
\mathscr{H} = zJ_c(\mathbf{S}_1\cdot\mathbf{S}_2) - K(S_{1,y}^2+S_{2,y}^2) - g\mu_{\rm B}(\mathbf{S}_1 + \mathbf{S}_2)\cdot \mathbf{H},
\end{equation}
where $\mathbf{S}_1$ and $\mathbf{S}_2$ are the spins representing the FM layers and $J_c$ and $z=2$ are the interlayer AFM exchange interaction and the coordination number. $K$ denotes the anisotropy constant for the uniaxial anisotropy along the $b$ direction. $g$, $\mu_{\rm B}$, $\mathbf{H}$ are the electron Laud\'e $g$ factor, Bohr magneton and magnetic field. Within the mean field description,  
the effective anisotropy field $H_{\rm A}$ and exchange field $H_{\rm E}$ are defined 
as 
\begin{equation}
    H_{\rm A} = \frac{2KS}{g\mu_{\rm B}}, \quad H_{\rm E}=\frac{zJ_cS}{g\mu_{\rm B}},
\end{equation}
where $S=3/2$ for Cr$^{3+}$. 
Define the spin magnetic moment $\mathbf{m}_i=g\mu_{\rm B}\mathbf{S}_i$ and the saturation magnetic moment $M_s=g\mu_{\rm B}S$, the Hamiltonian can be reformulated as 
\begin{equation}
\mathscr{H} = M_sH_E\cos(\theta_1-\theta_2) - 1/2M_sH_A\left( \cos^2{\theta_1} + \cos^2{\theta_2}\right) - M_s\mathbf{H}(\mathbf{\hat{m}_1} + \mathbf{\hat{m}_2}), \label{eq_simulated_M}
\end{equation}
where $\theta_1$ and $\theta_2$ are the angles between the positive $b$ direction and the spin magnetic moments along unit vectors $\mathbf{\hat{m}_1}$ and $\mathbf{\hat{m}_2}$. Fig.~\ref{FigS2} presents the simulated magnetization by minimizing Equation~\ref{eq_simulated_M} using the parameters tabulated in Table~1 in the main text, in good agreement with the experimental data shown in Fig. 1(c). The simulation reveals that for $\mathbf{H} \parallel a$, the expression $\theta_1+\theta_2=\pi$ holds true when the $\mathscr{H}$ is minimized. With this constraint enforced, an analytical expression, $\sin(\theta)=H/(2H_{\rm E}+H_{\rm A})$ describes the field evolution of the spin canting angle $\theta$. The simulated $\theta$ at $T$ = 2~K and 10~K as functions of magnetic field along the $a$ direction are plotted in Fig.~4(b) in the main text. 

\clearpage

\section{Structural domain volume fraction determination}
\begin{figure*}[htp]
  \includegraphics[width=0.8\linewidth]{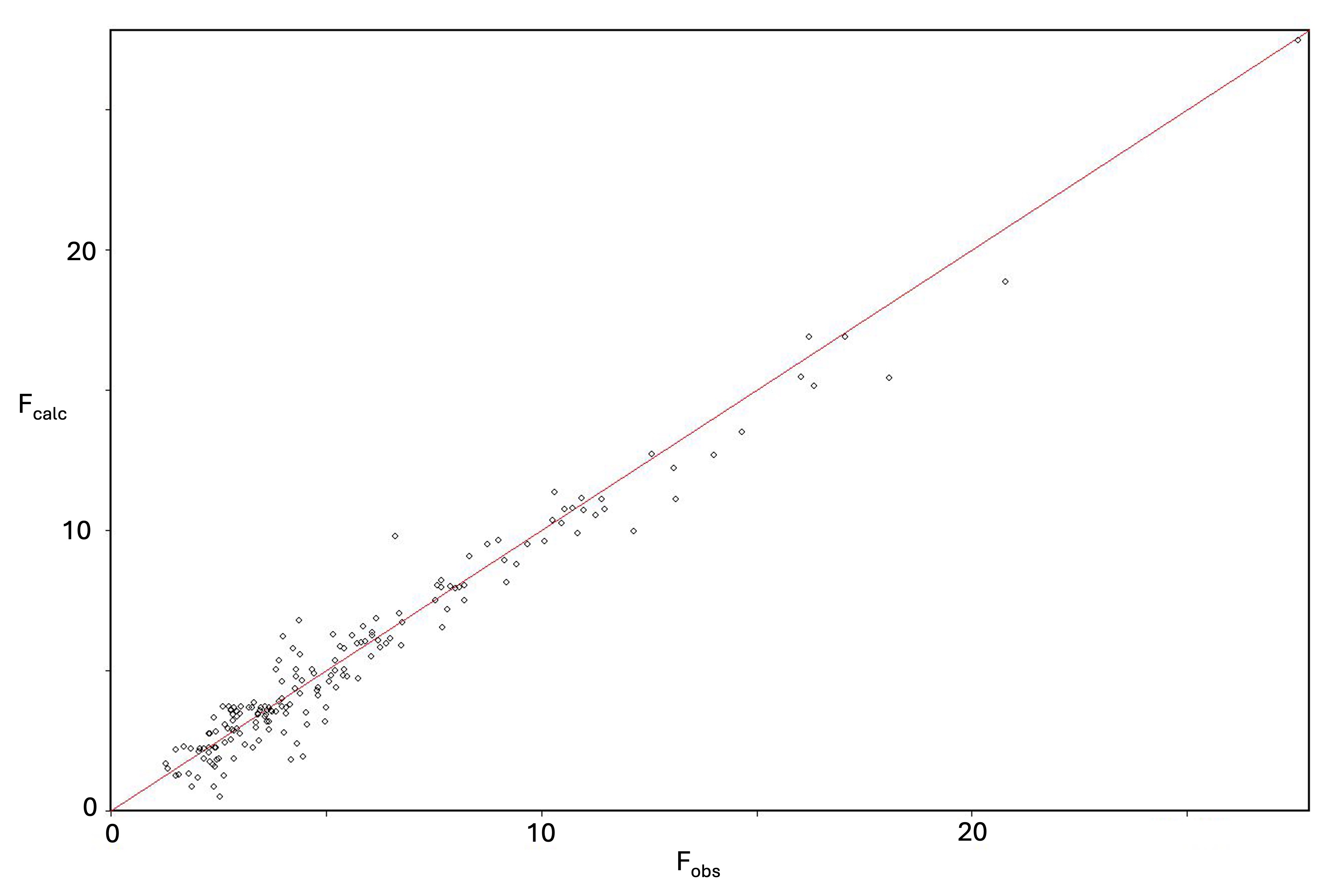}
  \caption{F$_{\rm calc}$ vs F$_{\rm obs}$ at $T$ = 165 K for the integer nuclear peaks used for the determination of the domain volume fractions.}
  \label{FigS3}
\end{figure*}

The monoclinic structure of CCPS can be transformed into a pseudo-hexagonal structure via
\begin{equation}
(a_{ph}, b_{ph}, c_{ph}) = (a_m, b_m, c_m) \mathbf{P}, 
\end{equation}
where the subscript $h$ and $m$ denote for $pseudo$-$hexagonal$ and $monoclinic$, and the transformation matrix
\begin{equation}
\mathbf{P} =
\begin{bmatrix}
1 & -1/2 & 2\\
0 & 1/2 & 0\\
0 & 0   & 3\\
\end{bmatrix}
\end{equation}
Accordingly, a given reflection ($H_m, K_m, L_m$) can be transformed to ($H_{ph}, K_{ph}, L_{ph}$) via
\begin{equation}
    (H_{ph}, K_{ph}, L_{ph})^T = \mathbf{P}^T (H_m, K_m, L_m)^T. \label{eq:m_to_h}
\end{equation}
A rotation operation converts the reflection via
\begin{equation}
    (H^\prime_{ph}, K^\prime_{ph}, L^\prime_{ph})^T = \mathbf{R} (H_{ph}, K_{ph}, L_{ph})^T, \label{eq:h_to_h}
\end{equation}
where $\mathbf{R}$ is the rotation matrix in the pseudo-hexagonal structure.  
 Putting Equation~(\ref{eq:m_to_h}) and (\ref{eq:h_to_h}) together and transforming the reflection back to the monoclinic frame, a reflection arising from a rotated structural domain can be indexed in the main domain frame via
\begin{equation}
    (H^\prime_m, K^\prime_m, L^\prime_m)^T = \mathbf{U} (H_m, K_m, L_m)^T, 
\end{equation}
with $\mathbf{U}$ being the effective transformation matrix given by
\begin{equation}
    \mathbf{U} ={\rm inv}(\mathbf{P}^T)\mathbf{R}\mathbf{P}^T.
\end{equation}
A reflection $(H^\prime_m, K^\prime_m, L^\prime_m)^T$ of the rotated domain could locate at fractional positions or integer positions, of which, the latter overlaps with a reflection of the main domain. Therefore, the domain volume fractions can be solved by reproducing the intensity of integer peaks. 
Neutron scattering intensities of a group of integer nuclear peaks collected at $T = $165 K was used for refining the domain volume fractions with the atomic positions and the displacement parameters refined from single crystal x-ray data collected at the same temperature [Table~\ref{tab:165K}]. Table~\ref{tab_domain_fraction} presents $\mathbf{U}$ of the rotated domains and the corresponding volume fractions refined using \textsc{Jana2020}~\cite{Jana2020}. The goodness-of-fit is shown in Fig.~\ref{FigS3}.

\begin{table*}[hb]
\caption{The atomic positions and displacement parameters in space group $C2/c$ for CCPS at $T$ = 165 K refined from single crystal x-ray data and used in the determination of domain volume fractions. All quantities are in fractional coordinates.}
\begin{ruledtabular}
\begin{tabular}{cccccccccc} 
  &$x$&$y$&$z$&$U_{11}$&$U_{22}$&$U_{33}$&$U_{12}$&$U_{13}$&$U_{23}$\\
  \hline
S&0.75165(8)&0.18486(5)&0.62578(3)&0.0049(2)&0.0077(2)&0.0082(2)&-0.0025(1)&0.0032(2)&-0.0042(1)\\
S&0.73167(8)&0.17470(4)&0.12569(3)&0.0061(2)&0.0065(2)&0.0072(2)&-0.0026(1)&0.0029(2)&-0.0031(1)\\
S&0.78025(8)&0.49532(4)&0.62364(4)&0.0050(2)&0.0037(2)&0.0087(2)&-0.0008(1)&-0.0027(2)&0.0006(1)\\
P&-0.05303(8)&0.33185(4)&0.16570(4)&0.0040(2)&0.0044(2)&0.0041(2)&0.0003(1)&0.0002(2)&0.0002(1)\\
Cr&0        &0.66466(3)&0.25      &0.0028(2)&0.0034(2)&0.0037(2)&0           &-0.0001(1)&0\\
Cu&-0.0652(2)&0.00221(7)&0.6454(1)&0.0101(2)&0.0064(2)&0.0611(4)&0.0000(1)&0.0169(2)&0.0002(2)\\
\end{tabular}
\end{ruledtabular}
\label{tab:165K}
\end{table*}

\begin{table}[hb]
\caption{The transformation matrices $\mathbf{U}$ from the main domain to rotated domains and the corresponding refined volume fractions.}
\begin{ruledtabular}
\begin{tabular}{c|ccc|ccc|ccc|ccc|ccc|ccc}

&\multicolumn{3}{c|}{Main domain}&\multicolumn{3}{c|}{60$^\circ$ domain}&\multicolumn{3}{c|}{120$^\circ$ domain}&\multicolumn{3}{c|}{180$^\circ$ domain}&\multicolumn{3}{c|}{240$^\circ$ domain}&\multicolumn{3}{c}{300$^\circ$ domain} \\
\hline
$\mathbf{U}$&1&0&0&1/2&-1/2&0&-1/2&-1/2&0&-1&0&0&-1/2&1/2&0&1/2&1/2&0 \\
&0&1&0&3/2&1/2&0&3/2&-1/2&0&0&-1&0&-3/2&-1/2&0&-3/2&1/2&0 \\
&0&0&1&1/3&1/3&1&1&1/3&1&4/3&0&1&1&-1/3&1&1/3&-1/3&1 \\
\hline
Volume fraction&\multicolumn{3}{c|}{0.424}      &\multicolumn{3}{c|}{0.003}            &\multicolumn{3}{c|}{0.054}             &\multicolumn{3}{c|}{0.461}             &\multicolumn{3}{c|}{0.020}             &\multicolumn{3}{c}{0.038}    
\end{tabular}
\end{ruledtabular}
\label{tab_domain_fraction}
\end{table}

\clearpage

\section{Magnetic structure refinement}

\begin{figure*}[htp]
  \includegraphics[width=0.6\linewidth]{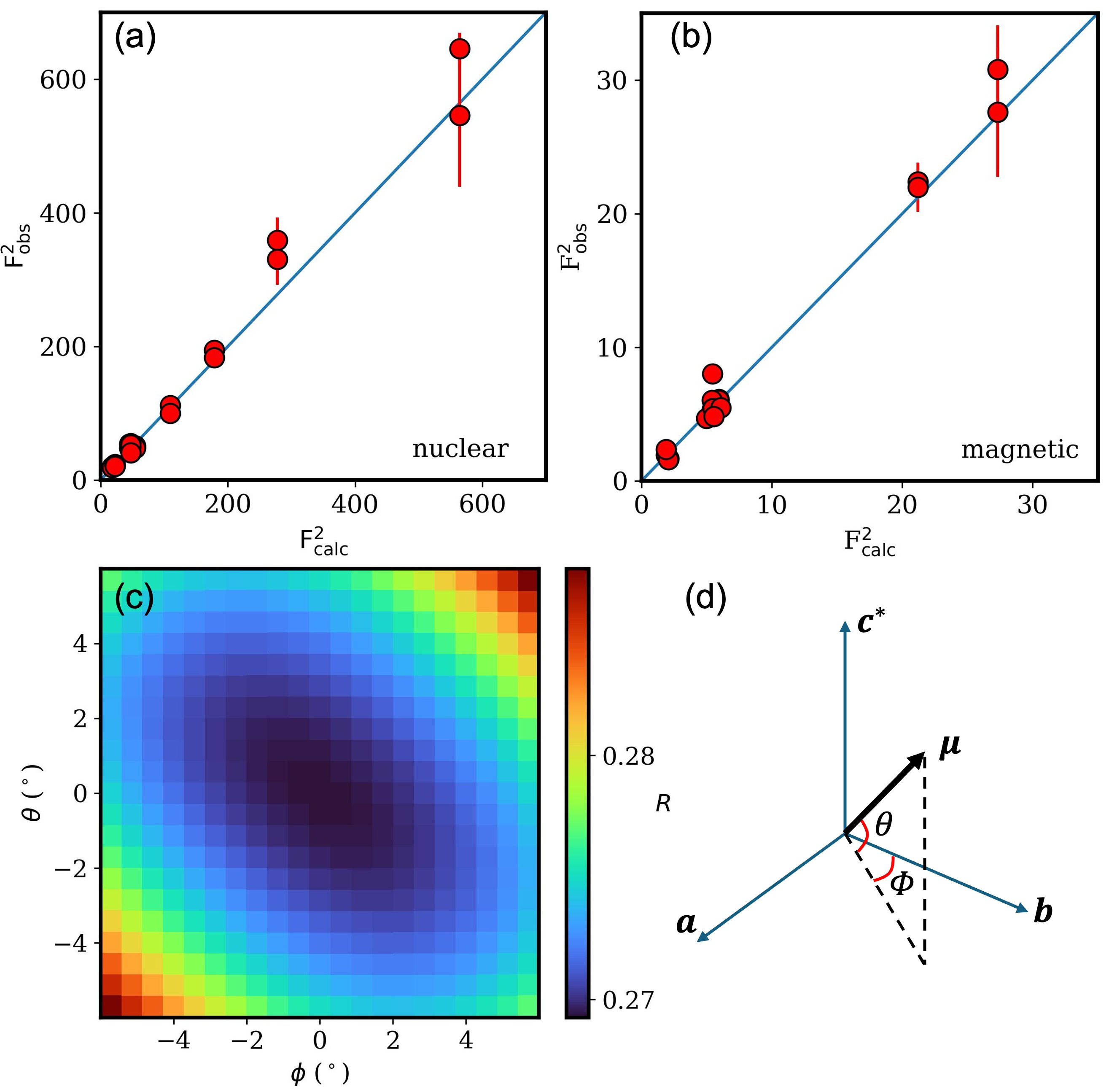}
  \caption{${F}^2_{\rm obs}~vs~{F}^2_{\rm cal}$ for the (a) nuclear and (b) magnetic peaks. (c) The residual $R$ as the moments cant away from the $b$ direction. (d) Illustration of the canting angle definition.}
  \label{FigS4}
\end{figure*}

The multi-domain magnetic structure refinement was carried out with an in-house developed python script. A group of integer nuclear peaks at $T$ = 5~K was identified to determine the scale factor (a factor of 0.875 was applied for the ($0$, $0$, $L$) peaks to correct for the misalignment of $\sim$ 1~g out of the total 8~g sample). For each peak, the criteria of determining the contributing domains is that they would assure integer peaks upon transforming the original peak to specific domains. Fig.~\ref{FigS4}(a) demonstrate the ${F}^2_{\rm obs}~vs~{F}^2_{\rm cal}$ for the scale factor determination where the atomic coordinates and displacement parameters are given in Table~\ref{tab:110K}, refined from single crystal x-ray data collected at $T$ = 110 K (lowest accessible $T$). Magnetic intensities were obtained by subtracting the $T$ = 5 K data from the $T$ = 50 K data for a list of integer peaks and the contributing domains determined following the same criteria. 

We first performed representational analysis was performed for the parent structure [space group $Pc$] with a propagation vector $\mathbf{k}=(0, 0, 0)$ using SARA\textit{h}~\cite{SARAh}. The magnetic representation can be decomposed into $\Gamma_{\rm mag}=3\Gamma^1_1+3\Gamma^1_2$, listed in Table~\ref{tab1}. $\Gamma_1$ corresponds to magnetic space group $Pc$, which allows for FM along the $b$ direction and A-type antiferromagnetism (AFM) along the $a$ and $c$ directions. On the contrary, $\Gamma_2$ corresponds to magnetic space group $Pc^\prime$, allowing FM along $a$ and $c$ directions and A-type AFM along the $b$ direction. A constrained refinement with collinear moments with equal size on the two independent Cr sites was then performed. The refined magnetic structure (AFM$_b$) is described by magnetic space group $Pc^\prime$ with A-type AFM along the $b$ direction and the ordered moment size $\mu$ = 3.03(5)~$\mu B$ (Fig.~\ref{FigS4}(b)). Next, within this magnetic space group, FM components along the $a$ and $c$ directions are introduced, which increases the residual,   
\begin{equation}
R = \frac{\sum_i{|F^2_{{\rm cal},i}-F^2_{{\rm obs},i}|}}{\sum_i{F^2_{{\rm obs},i}}}
\end{equation}
as shown in Fig.~\ref{FigS4}(c). In the above equation, $F^2_{\rm calc}$ and $F^2_{\rm obs}$ are calculated and observed peak intensities. Thus, within the resolution of the neutron diffraction data, we are convinced that the ground state ordered moments in CCPS align along the $b$ direction. Table~\ref{tab2} tabulates the observed and calculated intensities. For comparison, the calculated intensity of reported models in the literature are also included with $\mu$ = 3.03~$\mu_{\rm B}$.

\begin{table*}[hb]
\caption{The atomic positions and displacement parameters in space group $Pc$ for CCPS at $T$ = 110 K refined from single crystal x-ray data and used in the magnetic structure refinement. All quantities are in fractional coordinates.}
\begin{ruledtabular}
\begin{tabular}{cccccccccc} 
  &$x$&$y$&$z$&$U_{11}$&$U_{22}$&$U_{33}$&$U_{12}$&$U_{13}$&$U_{23}$\\
  \hline
Cr&0.1304(2)&-0.1651(2)&-0.6438(1)&0.0066(5)&0.0077(7)&0.0096(5)&-0.0030(5)&0.0006(4)&-0.0009(5)\\
Cr&-0.3724(2)&-0.6645(2)&-0.6427(1)&0.0082(5)&0.0078(7)&0.0076(5)&-0.0030(5)&0.0014(4)&-0.0009(5)\\
Cu&-0.3033(2)&-0.9982(1)&-0.5363(1)&0.0124(5)&0.0108(6)&0.0422(8)&-0.0009(5)&0.0127(5)&0.0002(6)\\
Cu&0.0611(2)&-0.4982(1)&-0.7509(1)&0.0128(5)&0.0102(6)&0.0378(7)&-0.0007(5)&0.0080(5)&0.0011(6)\\
S&-0.1284(4)&-0.3078(3)&-0.7720(2)&0.0079(9)&0.008(1)&0.0083(9)&0.0009(8)&0.0016(7)&0.0014(9)\\
S&-0.1465(4)&-0.6806(3)&-0.7693(2)&0.0060(8)&0.008(1)&0.0074(8)&-0.0018(8)&0.0000(6)&-0.0001(8)\\
S&-0.6383(5)&-0.1693(3)&-0.7670(2)&0.0087(9)&0.008(1)&0.0104(9)&0.0004(8)&0.0035(7)&0.0021(8)\\
S&-0.6157(4)&-0.8214(3)&-0.7638(2)&0.0082(9)&0.010(1)&0.0096(9)&0.0001(8)&0.0016(7)&-0.0019(8)\\ 
S&-0.0945(4)&-0.0029(3)&-0.7642(2)&0.0098(7)&0.010(1)&0.0147(9)&-0.0008(8)&0.0001(7)&0.0000(8)\\
S&-0.5802(4)&-0.5076(3)&-0.7759(2)&0.0070(8)&0.012(1)&0.012(1)&0.0021(8)&0.0025(7)&0.0006(9)\\
P&-0.4324(4)&-0.3291(3)&-0.7300(2)&0.0119(9)&0.008(1)&0.008(1)&0.0016(9)&0.0033(8)&0.0002(8)\\
P&0.0823(4)&-0.8352(3)&-0.7265(2)&0.009(1)&0.013(1)&0.0046(9)&-0.0008(9)&0.0000(8)&0.0013(8)\\
S&-0.1403(4)&-0.5032(3)&-0.5233(2)&0.0067(7)&0.0057(9)&0.0078(9)&0.0010(7)&-0.0013(6)&-0.0009(8)\\
S&0.3914(5)&-0.6698(2)&-0.5203(2)&0.0102(9)&0.009(1)&0.0102(9)&-0.0030(8)&0.0019(8)&-0.0009(8)\\
S&0.3451(4)&-0.0063(3)&-0.5127(2)&0.0075(7)&0.0054(9)&0.0095(8)&0.0016(7)&-0.0005(6)&0.0011(7)\\
S&0.3688(4)&-0.3221(3)&-0.5233(2)&0.0084(9)&0.012(1)&0.012(1)&0.0029(8)&0.0043(8)&0.0026(9)\\
S&-0.1181(4)&-0.8092(3)&-0.5151(2)&0.0086(9)&0.009(1)&0.0121(9)&0.0028(8)&0.0032(7)&0.0037(9)\\
S&-0.0975(5)&-0.1809(3)&-0.5176(2)&0.0099(9)&0.010(1)&0.0124(9)&-0.0024(8)&0.0036(7)&-0.0026(8)\\
P&-0.3253(4)&-0.3357(2)&-0.5618(2)&0.0062(8)&0.004(1)&0.012(1)&-0.0004(8)&0.0018(8)&0.0010(8)\\
P&0.1876(4)&-0.8283(3)&-0.5581(2)&0.0034(9)&0.008(1)&0.0096(9)&0.0015(8)&-0.0008(8)&0.0002(8)\\

\end{tabular}
\end{ruledtabular}
\label{tab:110K}
\end{table*}

\begin{table*}[hb]
\caption{Basis vectors (BVs) for the space group $Pc$ with magnetic propagation vector $\mathbf{k}=(0,0,0)$. The decomposition of the magnetic representation is $\Gamma_{\rm mag}=3\Gamma^1_1+3\Gamma^1_2$. The two magnetic Cr atoms sites locate at Wyckoff position $2a$ with fractional coordinates: Cr1:(0.14724, 0.33531, 0.39605), Cr3:(0.64443, 0.83510, 0.39721) with Cr2 and Cr4 generated by the $c$-gliding operation (X, -Y+1, Z+1/2), respectively.}
\begin{ruledtabular}
\begin{tabular}{cccccccccccc} 
&&&\multicolumn{3}{c}{component}&&&&\multicolumn{3}{c}{component}\\
\hline
IR&BV&Atom&$m_a$&$m_b$&$m_c$&IR&BV&Atom&$m_a$&$m_b$&$m_c$\\
$\Gamma_1$&\multirow{2}{*}{$\psi_1$} &Cr1 &1&0&0     &$\Gamma_2$    &\multirow{2}{*}{$\psi_1$}&Cr1&1&0&0\\ 
($Pc$)    &                          &Cr2 &-1&0&0    &($Pc^\prime$) &                         &Cr2&1&0&0\\
          &\multirow{2}{*}{$\psi_2$} &Cr1 &0&1&0     &              &\multirow{2}{*}{$\psi_2$}&Cr1&0&1&0\\
          &                          &Cr2 &0&1&0     &              &                         &Cr2&0&-1&0\\
          &\multirow{2}{*}{$\psi_3$} &Cr1 &0&0&1     &              &\multirow{2}{*}{$\psi_3$}&Cr1&0&0&1\\
          &                          &Cr2 &0&0&-1    &              &                         &Cr2&0&0&1\\
          &\multirow{2}{*}{$\psi_4$} &Cr3 &1&0&0     &              &\multirow{2}{*}{$\psi_4$}&Cr3&1&0&0\\ 
          &                          &Cr4 &-1&0&0    &              &                         &Cr4&1&0&0\\
          &\multirow{2}{*}{$\psi_5$} &Cr3 &0&1&0     &              &\multirow{2}{*}{$\psi_5$}&Cr3&0&1&0\\ 
          &                          &Cr4 &0&1&0     &              &                         &Cr4&0&-1&0\\
          &\multirow{2}{*}{$\psi_6$} &Cr3 &0&0&1     &              &\multirow{2}{*}{$\psi_6$}&Cr3&0&0&1\\ 
          &                          &Cr4 &0&0&-1    &              &                         &Cr4&0&0&1\\
\end{tabular}
\end{ruledtabular}
\label{tab1}
\end{table*}

\begin{table}[bh]
\caption{Comparison of observed and calculated magnetic intensities for three A-type AFM models with moments along $b$ (AFM$_b$, this work), $a$ (AFM$_a$~\cite{Park2022, Wang2023}) and $ab$ plane diagonal (AFM$_{ab}$~\cite{Cajipea1996}) directions. Calculation of the residual $R$ excludes the non-observed (0, 2, 0) peak.}
\begin{ruledtabular}
\begin{tabular}{ccccc}
%\hline
%\hline
Reflection&Observation&AFM$_b$ &AFM$_a$ &AFM$_{ab}$ \\
\hline
(0, 0, 1)&31 (3)&27.30&27.30&27.30 \\
(0, 0, -1)&28 (5)&27.30&27.30&27.30 \\
(0, 0, 3)&22 (1)&21.20&21.20&21.20 \\
(0, 0, -3)&22 (2)&21.20&21.20&21.20 \\
(1, 1, 0)&5 (0)&5.40&2.09&2.13 \\
(1, -1, 0)&6 (0)&5.95&2.31&7.32 \\
(-1, 1, 0)&6 (0)&5.95&2.31&7.32 \\
(-1, -1, 0)&6 (1)&5.40&2.09&2.13 \\
(1, 1, 1)&2 (0)&1.89&1.09&1.10 \\
(1, -1, 1)&2 (0)&2.08&1.21&2.42 \\
(0, 2, 2)&5 (1)&5.46&13.91&7.57 \\
(0, 2, -2)&5 (1)&5.46&13.91&7.57 \\
(0, -2, 2)&5 (1)&5.46&13.91&7.57 \\
(0, -2, -2)&8 (1)&5.46&13.91&7.57 \\ 
(1, 1, 2)&5 (1)&4.97&3.65&3.66 \\
(1, -1, 2)&5 (1)&5.49&4.03&6.05 \\ 
(-1, 1, 2)&5 (1)&6.09&3.18&7.19 \\
(-1, -1, 2)&5 (1)&5.54&2.89&2.92 \\
(-1, 1, 3)&2 (0)&2.07&1.44&2.31 \\
(-1, -1, 3)&2 (1)&1.89&1.31&1.32 \\
(0, 2, 0)&0 (0)&0.00&15.77&3.94\\
\hline
$R$      &           &0.07 &0.34 &0.17\\
%\hline
%\hline
\end{tabular}
\end{ruledtabular}
\label{tab2}
\end{table}

\clearpage

\section{Extracting spin canting angle from fitting magnetic peak intensity}
\begin{figure*}[htp]
  \includegraphics[width=0.4\linewidth]{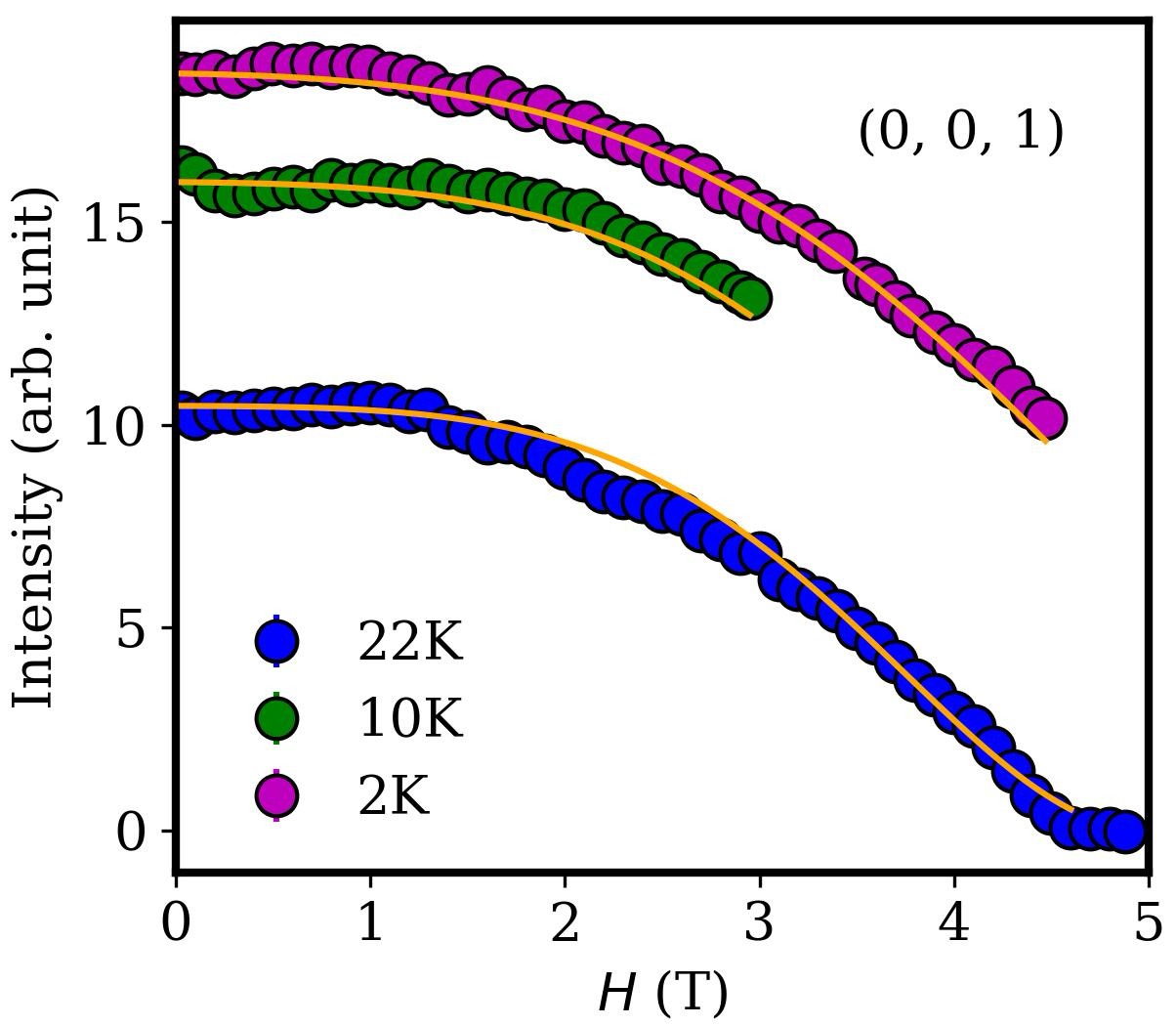}
  \caption{Field evolution of the magnetic peak (0, 0, 1) intensity at various temperatures. Solid lines are empirical model fits based on Eq.~\ref{fitting}.}
  \label{FigS5}
\end{figure*}

In the simplified model considering rigid FM layers AFM-coupled along the layer normal, an empirical model~\cite{Ye2022} 

\begin{equation}
\theta \propto AH^2+BH+C, \label{fitting}
\end{equation}
can be used to correlate the canting angle $\theta$ between the FM moment and the $b$ direction with the magnetic field strength along the $a$ direction, where $A$, $B$, $C$ are constants. Magnetic peak intensity is subsequently determined from the canting angle. Fig.~\ref{FigS5} presents the field evolution of the magnetic peak (0, 0, 1) intensity at various temperatures [$T$ = 27 K in Fig.~3(a)], simultaneously fitted with the empirical model, which allows us to extracted the canting angle as a function of field, plotted in Fig.~3(b). 

\clearpage

\bibliographystyle{h-physrev}
\bibliography{reference.bib}